\documentclass[final,5p,times,twocolumn]{elsarticle}

\usepackage{graphicx}
\usepackage{amssymb}
\usepackage{mathrsfs}

\usepackage{lineno}
\usepackage{listings}

\usepackage{amsmath,amsfonts,amsthm} 
\usepackage{xcolor}
\usepackage{enumitem}
\usepackage{booktabs}
\usepackage{soul}
\usepackage{ulem}
\usepackage{threeparttable}
\usepackage{hyperref}

\definecolor{mygreen}{rgb}{0,0.6,0}
\definecolor{mygray}{rgb}{0.5,0.5,0.5}
\definecolor{mymauve}{rgb}{0.58,0,0.82}
\biboptions{square}

\bibpunct{(}{)}{;}{a}{}{,} 

\newcommand{\pyedra}[0]{Pyedra}

\journal{astronomy \& computing}

\begin{document}

\begin{frontmatter}


\title{
    Easy asteroid phase curve fitting for the Python ecosystem: \pyedra~
}




\author[iate,famaf]{Milagros R. Colazo}
\author[cifasis,iate]{Juan B. Cabral}
\author[iate,oac]{Martín Chalela}
\author[duke]{Bruno O. S\'anchez}

\address[iate]{
   Instituto de Astronom\'ia Te\'orica y Experimental -
   Observatorio Astron\'omico de C\'ordoba (IATE, UNC--CONICET),
   C\'ordoba, Argentina.}
\address[famaf]{
	Facultad de Matem\'atica, Astronom\'{\i}a y F\'{\i}sica,
    Universidad Nacional de C\'ordoba (FaMAF--UNC)
	Bvd. Medina Allende s/n, Ciudad Universitaria,
    X5000HUA, C\'ordoba, Argentina}
\address[cifasis]{
   Centro Internacional Franco Argentino de Ciencias de la
   Informaci\'on y de Sistemas (CIFASIS, CONICET--UNR),
   Ocampo y Esmeralda, S2000EZP,
   Rosario, Argentina.}
\address[duke]{Department of Physics, Duke University, 120 Science Drive, Durham, NC, 27708, USA}

\begin{abstract}
A trending astronomical phenomenon to study is the
the variation in brightness of asteroids, caused by 
its rotation on its own axis, non-spherical shapes, 
changes of albedo along its surface and 
its position relative to the sun. 
The latter behavior can be visualized on a ``Phase Curve'' (phase angle vs. reduced magnitude). To enable the comparison between several models proposed for this curve we present a Python package called 
\pyedra.
%
\pyedra~implements three phase-curve-models,
and also providing capabilities for visualization as well as integration with external datasets.
%
%
The package is fully documented and tested following a strict quality-assurance workflow, whit a user-friendly programmatic interface.
%
%
In future versions, we will include more models, and additional estimation of quantities derived
from  parameters like diameter, and types of albedo; as well as enabling
correlation of information of physical and orbital parameters.

\end{abstract}

\end{frontmatter}



\section{Introduction}

The brightness variation of asteroids is a fascinating astronomical phenomenon to study. One cause of the object's varying magnitude is its rotation about its own axis. This is because asteroids have non-spherical shapes and albedo differences along their surface. On the other hand, the brightness of an asteroid will also vary just by moving in its orbit around the Sun. When the object is close to opposition, i.e. at angles close to $0^{\circ}$, sunlight hitting the asteroid and  light reflected from the object's surface will come from the same direction, causing the object to have a maximum in apparent brightness. As it moves in its orbit and its phase angle increases, the Sun's light will begin to cast shadows on the asteroid's surface causing a decrease in its brightness. In summary, the magnitude of an asteroid drops as it approaches the opposition and as it moves away the magnitude will begin to grow again. This behavior can be visualized on a phase angle ($\alpha$) vs. reduced magnitude $V$ diagram, known as "Phase Curve". Although several models have been proposed to describe this curve, there is no comprehensive tool available that provides with the necessary fitting procedures and enables a reliable comparison between these models.

In 1989, \citeauthor{1989aste.conf..524B} proposed the $G$ model (also known as $H, G$ model), a semi-empirical model derived from the basic principles of radiative transfer theory with some assumptions \citep{2015AJ....150...75W}. \cite{1996LPI....27.1193S} proposed an empirical tri-parametric phase function model valid for phase angles in the range of $0^{\circ}-40^{\circ}$. There is a third model for the phase function proposed by \citeauthor{2010Icar..209..542M} in 2010. It is a model similar to Bowell's but replaces the $G$ parameter with two parameters $G_1$ and $G_2$, making it also a tri-parametric model. According to \citeauthor{2010Icar..209..542M}, the $G$ model is a good approximation in the region of $10^{\circ}{\sim}60^{\circ}$, while the $G_1$ and $G_2$ model works well also for angles close to the opposition (${\sim}0^{\circ}$).

Many large sky surveys are currently in operation. Thousands of asteroids are observed by these telescopes, providing a unique opportunity to study them.
Some of these large sky surveys are Gaia \citep{2018A&A...616A..13G}, TESS \citep{2015JATIS...1a4003R}, in the near future the Vera Rubin's Legacy Survey of Space and Time (LSST, \citealp{Schwamb_2019}). This poses the opportunity to characterize the phase curve of a large numbers of asteroids. We must be prepared to take full advantage of this information. One of the main analysis with these datasets would be the calculation of the absolute magnitude $H$ of hundreds or thousands of these objects, enabling also the estimation of their diameters. The parameters $G$, $G_1$, $G_2$, $b$ can provide a good estimate of the albedo of the asteroids and, even more, can help in the taxonomic classification \citep{1996LPI....27.1193S,2000Icar..147...94B, 2019P&SS..169...15C}.

In this context of "big data for asteroids" we developed \pyedra~. Pyedra enables the analysis of large amounts of data, it provides the parameters of the selected phase function model that best fits the observations. 
Thus, it can quickly create parameter catalogs for large databases as well as providing the possibility of working with non-survey data, i.e. personal observations.

This paper is organized as follows: in Section 2 we provide a brief description of the algorithm. In Section 3 we introduce technical details about the Pyedra package. In Section 4 we present the conclusions and future perspectives. 

\section{The Algorithm}\label{algorithm}

In this section, we present the three phase function models implemented in Pyedra and for each one of them, we provide details on the method used for parameter estimation. In general we adopt the procedure proposed by \cite{2010Icar..209..542M}, hereafter M10.

\subsection{H, G model}\label{HG}
The $H, G$ phase function model for asteroids can be described analytically through the following equation \citep{2010Icar..209..542M}:
\begin{equation}
    V(\alpha)=H-2.5\log_{10}[(1-G)\Phi_{1}(\alpha)+G\Phi_{2}(\alpha)], 
\label{eqn:1}
\end{equation}
where $H$ and $G$ are the two free parameters of the model, $\alpha$ is the phase angle, $V(\alpha)$ is the reduced $V$ magnitude (brightness on Johnson's filter $V$ normalized at 1 AU from the Sun and the observer), $\Phi_{1}$($\alpha$) and $\Phi_{2}$($\alpha$) are two basis function normalized at unity for $\alpha=0^{\circ}$. The base functions can be accurately approximated by:
\begin{equation}
\begin{aligned}
    \Phi_{1}(\alpha) &= \exp & \left(-3.33\tan^{0.63}\frac{1}{2}\alpha \right ), \\
    \Phi_{2}(\alpha) &= \exp & \left(-1.87\tan^{1.22}\frac{1}{2}\alpha \right ).
\label{eqn:2}
\end{aligned}
\end{equation}
\noindent
 To obtain the value of $H$ and $G$, M10 proposes to write to the reduced magnitude as:
 
 \begin{equation}
    10^{-0.4V(\alpha)}=a_{1}\Phi_{1}(\alpha)+a_{2}\Phi_{2}(\alpha),
\label{eqn:3}
\end{equation}
\noindent
then we can write the absolute magnitude $H$ and the coefficient $G$ as:
\begin{equation}
\begin{aligned}
    H &= -2.5\ \log_{10}(a_{1}+a_{2}), \\
    G &= \frac{a_{2}}{a_{1}+a_{2}}.     
\end{aligned}
 \label{eqn:4}
\end{equation}
\noindent
The coefficients $a_1$ and $a_2$ are estimated from the observations using the standard method of least squares.

\subsection{\texorpdfstring{H, G$_1$, G$_2$ }\ model}\label{HG1G2}
This three parameter magnitude phase function can be described as in M10:
\begin{equation}
    \begin{split}
    V(\alpha) & = H-2.5\log_{10}[G_{1}\Phi_{1}(\alpha)+G_{2}\Phi_{2}(\alpha)\\
              & +(1-G_{1}-G_{2})\Phi_{3}(\alpha)],
    \label{eqn:5}
    \end{split}
\end{equation}

\noindent
where $\Phi_{1}$(0°)=$\Phi_{2}$(0°)=$\Phi_{3}$(0°)=1. $H$, $G_1$ and $G_2$ are the parameters of the model, $\alpha$ is the phase angle, $V(\alpha$) is the reduced magnitude and $\Phi_{1}$, $\Phi_{2}$, $\Phi_{3}$ are basis functions.

These basis are defined piecewise using linear terms as well as cubic splines \citep{2016P&SS..123..117P} along the orbit.

In this case we write the reduced magnitude as:
\begin{equation}
    10^{-0.4V(\alpha)}=a_{1}\Phi_{1}(\alpha)+a_{2}\Phi_{2}(\alpha)+a_{3}\Phi_{3}(\alpha).
\label{eqn:6}
\end{equation}
\noindent
For this calculation we use the tabulated values for the base functions presented in \cite{2016P&SS..123..117P}. The model free parameters can be obtained from:
\begin{equation}
\begin{aligned}
    H &= -2.5\ \log_{10}(a_{1}+a_{2}+a_{3}), \\
    G_{1} &= \frac{a_{1}}{a_{1}+a_{2}+a_{3}}, \\
    G_{2} & = \frac{a_{2}}{a_{1}+a_{2}+a_{3}}.    
\label{eqn:7}
\end{aligned}
\end{equation}
\noindent
The coefficients $a_1$, $a_2$ and $a_3$ are estimated from observations using the method of least squares.

\subsection{Shevchenko model}\label{Shev}
This model is described in the following equation \citep{1996LPI....27.1193S}:
\begin{equation}
    V(1,\alpha)=V(1,0)-\frac{a}{1+\alpha}+b\cdot\alpha ,
\label{eqn:8}
\end{equation}
\noindent
 where $a$ characterizes the amplitude of the so-called ``opposition effect'', $b$ is a parameter describing the linear term of the phase-magnitude relationship, $\alpha$ is the phase angle and $V(1,0)$ is the absolute magnitude.

Although M10 does not present this model in its work, we have extended the use of least squares for this case as well given that Shevchenko's formula is already written in the form of a linear equation.

\section{Technical details about the Pyedra package}

\subsection{User functionalities and application example}
The Pyedra package consists of 3 main functions to perform the fitting of observations. Each of these functions corresponds to one of the models mentioned in Section \ref{algorithm}. These functions are:
\begin{itemize}
    \item \texttt{HG\_fit()}: fits the \ref{HG} model to the observations.
    \item \texttt{HG1G2\_fit()}: fists the \ref{HG1G2} model to the observations.
    \item \texttt{Shev\_fit()}: fits the \ref{Shev} model to the observations.
\end{itemize}

All these functions return an object that we call \texttt{PyedraFitDataFrame}, containing the parameters obtained after the fit with a format that is analogous to a pandas Dataframe. The \texttt{plot()} method of this object returns:
\begin{itemize}
    \item a graph of our observations in the plane ($\alpha$, $V$) together with the fitted model.
    \item any of the graphics that pandas allows to create.
\end{itemize}

Pyedra also offers the possibility to add Gaia observations to the user's sample. This is done by using:
\begin{enumerate}
    \item \texttt{.load\_gaia()} to read the files containing Gaia observations.
    \item \texttt{.merge\_obs()} to merge the user and Gaia tables.
    \item One can apply any of the above functions to this new dataframe.
\end{enumerate}

This is a very interesting feature because, in general, observations from the ground correspond to small phase angles. In contrast, Gaia can only observe for phase angles $\alpha > 10$° \citep{2018A&A...616A..13G}. Both sets of data are complementary, thus achieving a more complete coverage of the phase angle space. This also leads to a better determination of the phase function parameters.

\begin{figure}[ht!]
    \centering
    \includegraphics[width=1\columnwidth]{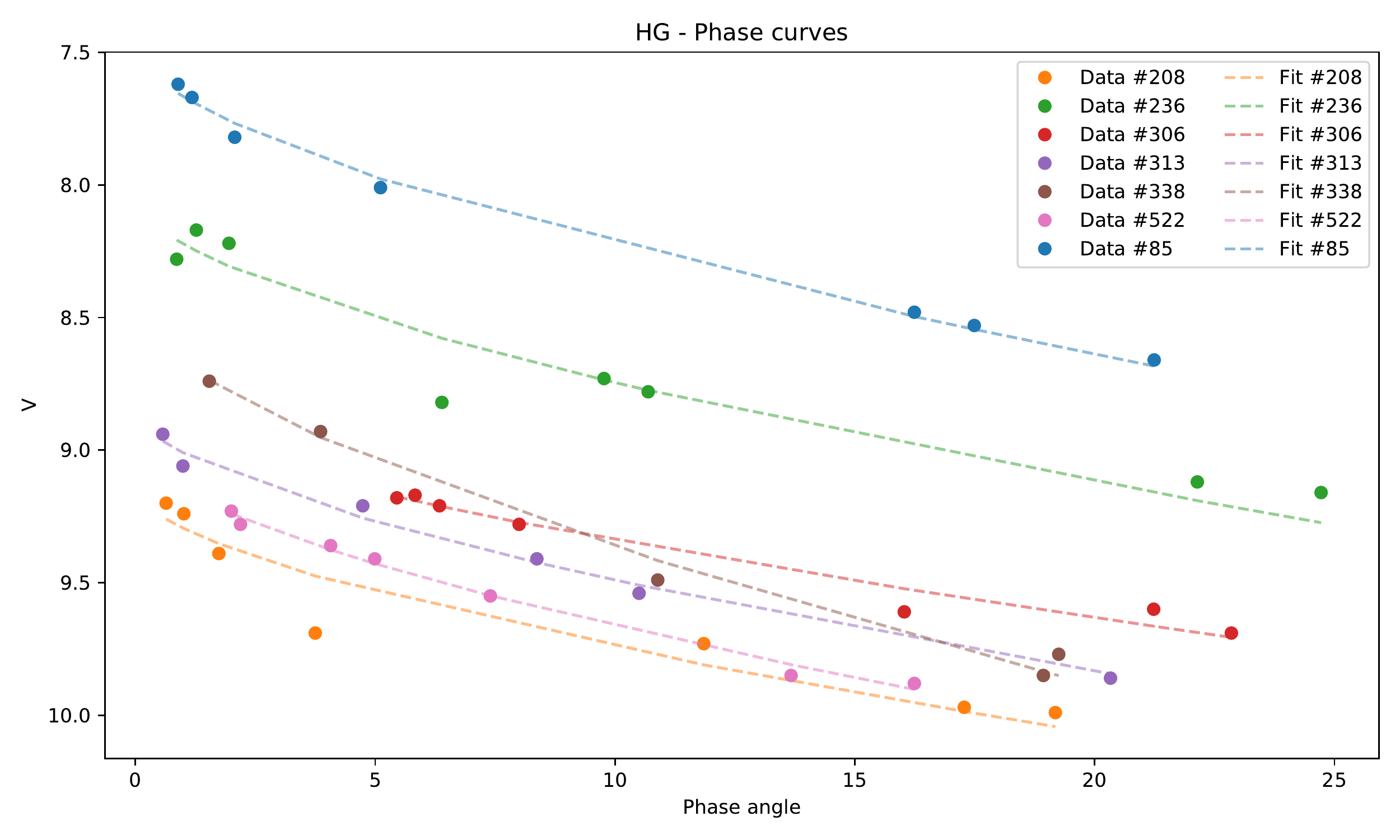}
    \caption{Example of the figure obtained for the model $H, G$. The points correspond to the observations and the dotted lines to the best fit. Points of the same color correspond to the same object. }
    \label{fig:fig1}
\end{figure}

\begin{figure}[ht!]
    \centering
    \includegraphics[width=1\columnwidth]{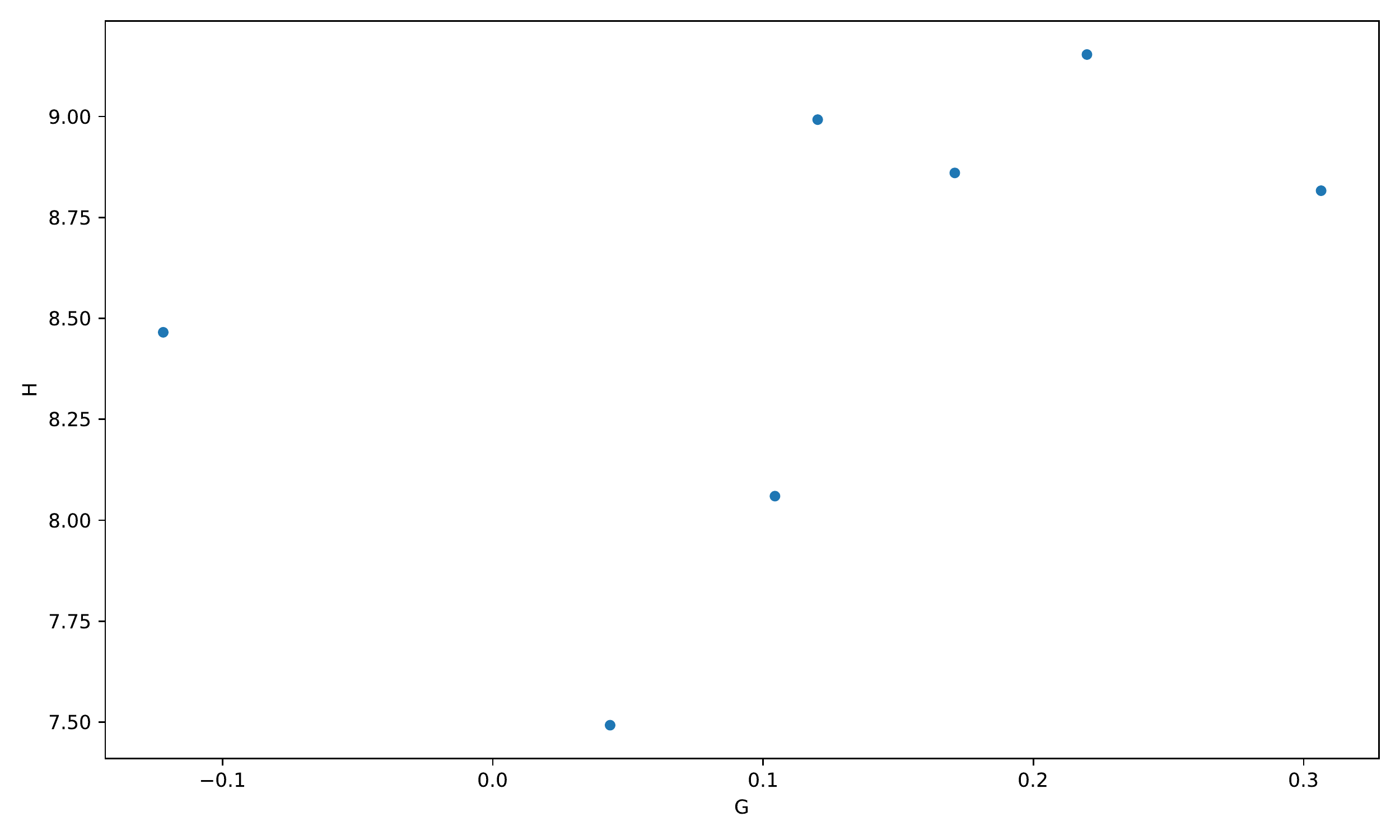}
    \caption{Example of the 'scatter' graphic of pandas dataframe, also available for PyedraDataFrame.}
    \label{fig:fig2}
\end{figure}

As a simple usage application we show how to calculate the parameters of the $H, G$ model and how to plot the observations with the fit obtained using the dataset of \cite{2019P&SS..169...15C}. The respective plots are shown in Figs.~\ref{fig:fig1}~\&~\ref{fig:fig2}. This dataset is provided with Pyedra for the user to test the functionalities. 

\begin{lstlisting}[language=Python]
>>> import pyedra
>>> import pandas as pd
>>> import matplotlib.pyplot as plt

# load the data
>>> df = pyedra.datasets.load_carbognani2019()

# fit the data
>>> HG = pyedra.HG_fit(df)
    id         H   error_H         G   error_G         R
0   85  7.492423  0.070257  0.043400  0.035114  0.991422
1  208  9.153433  0.217270  0.219822  0.097057  0.899388
2  236  8.059719  0.202373  0.104392  0.094382  0.914150
3  306  8.816185  0.122374  0.306459  0.048506  0.970628
4  313  8.860208  0.098102  0.170928  0.044624  0.982924
5  338  8.465495  0.087252 -0.121937  0.048183  0.992949
6  522  8.992164  0.063690  0.120200  0.028878  0.991757
PyedraFitDataFrame - 7 rows x 6 columns

# take the mean value of H
>>> HG.H.mean()
>>> 8.54851801238607

# plot the data and the fit
>>> HG.plot(df=df, ax=None)
>>> plt.show()
<AxesSubplot:title={'center':'Phase curves'}, 
xlabel='Phase angle', ylabel='V'>

# scatter plot of G vs H
>>> HG.plot(x='G', y='H', kind='scatter')
>>> plt.show()
<AxesSubplot:xlabel='G', ylabel='H'>

\end{lstlisting}

\subsection{Quality assurance}

Software quality assurance refers to the set of standards and procedures that must be used in order to verify that the software meets certain subjective quality criteria. The most common procedures to carry out this task are \textit{unit-testing} and \textit{code-coverage}. 

The purpose of \textit{unit-testing} is to check that each of the individual components of the software works as expected \citep{jazayeri2007some}. That is, we isolate a function from our code and verify that it works correctly. On the other hand, \textit{code-coverage} is a measure of how much of our software has been tested \citep{miller1963systematic}. In this way, we can identify parts of the code that we have not verified. In the Pyedra package we provide five suites of unit-tests that evaluate different sections of the code, reaching 99$\%$ of code-coverage. The testing suites are tested for Python versions 3.7, 3.8 and 3.9. We are also interested in the maintainability of Pyedra, therefore we have adopted PEP 8 – Style Guide for Python Code \citep{van2001pep} in such a way that our project meets current code standard and readability. For this purpose, we use the \textit{flake8}\footnote{\url{https://flake8.pycqa.org/en/latest/}} tool hat automatically detects any case where we are not respecting the style imposed by PEP 8 as well as programming errors, such as: "library imported but unused".

Finally, the entire source code is MIT-licensed and available in a public repository\footnote{\url{https://github.com/milicolazo/Pyedra}}. All changes and new versions of the package committed to this repository are automatically tested with continuous-integration services \footnote{\url{https://travis-ci.com/milicolazo/Pyedra}}$^{,}$ \footnote{\url{https://github.com/milicolazo/Pyedra/actions}}. Documentation is automatically generated from Pyedra docstrings and made public in the read-the-docs service\footnote{\url{https://pyedra.readthedocs.io/en/latest/?badge=latest}}.

Finally, Pyedra is available for installation on the PythonPackage-Index (PyPI)\footnote{\url{https://pypi.org/project/Pyedra/}}; 
and is currently going through registration process to appear in the Astrophysics Source Code Library (ASCL.net, \citealp{grosbol2010making})

\subsection{Integration with the Python scientific–stack}

Python has become an important programming language within the astronomical community \citep{2020JOSS....5.2732S}. This is mainly because it is a simple to use, free and versatile language for manipulating and visualizing data \citep{2012JCIS....3E...1F}.

Pyedra is built on top of the Python scientific stack: \textit{Pandas} \citep{mckinney2010data} since the main object on which Pyedra operates is a dataframe; \textit{Scipy} \citep{virtanen2020scipy} for function interpolation and fit of least squares optimization; \textit{Numpy} \citep{walt2011numpy} to manipulate arrays;  \textit{Matplotlib} \citep{hunter2007matplotlib} for the data visualization; and attrs\footnote{\url{https://www.attrs.org}} to facilitate the implementation of classes.

\subsubsection{Short comparison with other similar packages}

Pyedra's main objective is to calculate the parameters of different phase function models for large and small volumes of data. The \texttt{sbpy}\footnote{\url{https://sbpy.org/}} \citep{mommert2019sbpy} package offers the possibility to model phase curves. In this subsection, we will make a brief contrast between both projects.

Regarding the available models, \texttt{sbpy} and \pyedra~ share the HG and HG1G2 model. In the case of \texttt{sbpy}, the models HG12 (Revised H, G12 model by \citeauthor{2016P&SS..123..117P}) and a linear model are also available. Although these models were not considered in \pyedra~ (but will be implemented in the next release), we have included Shevchenko's model which is not present in \texttt{sbpy}.

On the other hand, \texttt{sbpy} does not provide the functionality to estimate the best fit model parameters (as \pyedra~ does) but returns other quantities derived from these parameters. In addition, \pyedra~ has an error estimate for each calculated value, something that is not present in \texttt{sbpy}.

Finally, \pyedra~'s main strength against \texttt{sbpy} is its simplicity of use. With sbpy we have not found a quick way to get phase function parameter catalogs for databases with large numbers of entries. With \pyedra~, the user can accomplish this task by just writing one line of code. The same is true for graphic capabilities: since plotting phase functions is one of \pyedra~'s features, one single method call allows to obtain a visualization of the phase function. It is also worth noticing that with \pyedra~ not only phase curve plots can be easily obtained, all pandas visualization tools are also available enabling a more comprehensive analysis of the resulting catalog. Moreover, as it is based on pandas' dataframe manipulation, the output catalog is simple to visualize, modify and to carry out different calculations from it.


\section{Conclusions}
In this paper, we present Pyedra, a python implementation for asteroid phase curve fitting. This package allows the user to fit three different models of phase functions to observations of asteroid phase angle and photometry.

Pyedra is suitable for analysis of private datasets, of one or more asteroids, as well as large volumes of information from any public survey data release such as TESS, Gaia, K2, among others. SiConsequently Pyedra is a tool that will enable the creation of phase curve model parameter catalogs for hundreds of thousands of asteroids.

Pyedra also offers the possibility of producing numerous visualization plots. Not only it can produce a graph of the phase functions but it makes available all the graphs natively offered for pandas dataframes. In this way, we provide the possibility of a complete analysis of the results obtained. 

As we have already mentioned, we are living in an era of big surveys. We must be able to have tools capable of processing the vast amount of data that these surveys constantly provide to the scientific community.

\subsection{Future Work}
Pyedra is still in development process, so there are still topics to be improved. 

The first thing to consider would be to have more phase function models added to those already offered by the package. In addition, it would be convenient to be able to estimate certain quantities derived from the parameters obtained, such as the diameter, the integral function, the different types of albedo, etc. It would be interesting to have a tool that (in the case of a database containing several asteroids) allows combining information on physical parameters with orbital parameters. For example, the possibility of studying which $G$ values the asteroids have for different semi-axes $a$.

Finally, we intend to add more large survey data for the user to combine with their observations, such as TESS, SDSS, etc. 

\section{Acknowledgments}
The authors would like to thank their families and friends, as well as the IATE astronomers for useful comments and suggestions.

This work was partially supported by the Consejo Nacional
de Investigaciones Cient\'ificas y T\'ecnicas (CONICET, Argentina).
M.R.C., J.B.C and M.Ch. were supported by a fellowship from CONICET.

This research employed the
http://adsabs.harvard.edu/, Cornell University xxx.arxiv.org repository,
the Python programming language, the Numpy and Scipy libraries,
and the other packages utilized can be found at the 
GitHub webpage for Pyedra.


\bibliographystyle{aa}
\bibliography{main.bib}

\end{document}